# A Reliable Millimeter-Wave Quadrature Interferometer

M. Gilmore, W. Gekelman, K. Reiling, and W.A. Peebles
*Institute of Plasma and Fusion Research, University of California, Los Angeles, California 90024*

A simple, low-cost millimeter-wave (70 GHz) interferometer with a phase accuracy better than ± 2 degrees, and a response time of 10 ns is described. The simplicity of this interferometer makes it ideal for measurement of electron densities in laboratory or semiconductor processing plasmas. The relative high accuracy and low cost are attained by a homodyne system with a quadrature intermediate frequency (IF) mixer, now commercially available in millimeter-wave frequency bands. The design and construction of magnetic shielding for the system isolator, which may be required around magnetically-confined plasmas, is also described.

## 1. Introduction

Although Langmuir probes are able to measure normalized density profiles, they are unreliable for determining absolute density in the presence of strong (B > 500 Gauss) magnetic fields. Comparison of the density from either the electron or ion saturation currents can be a factor of three lower than densities derived from wave dispersion. In applications where the absolute density needs to be known an alternative technique must be employed. The measurement of wave dispersion is time consuming, however microwave interferometry can be an attractive alternative.

Many interferometry applications, such as the measurement of electron densities $n_e$ = $10^{17}$-$10^{18}$ $m^{-3}$ in plasmas ranging in size from a few centimeters (cm) to many tens of cm, are best handled at millimeter-wave frequencies. Typically, millimeter-wave interferometers have been either rather low resolution homodyne systems, more complicated frequency modulated homodyne systems [1], or high resolution heterodyne systems which are relatively expensive to construct [2]. In unmodulated homodyne systems utilizing a mixer with a single IF output, the output voltage is $V_{IF}$ = Acos(theta) Here A is the amplitude and q is the phase angle between the radio frequency (RF) and local oscillator (LO) inputs. Generally $V_{IF}$ is a function of both phase angle and RF power. Thus, there is ambiguity between amplitude and phase changes, and the user must resort to counting the number of 2 pi phase shifts (fringes), and estimating fractions of a fringe. Additionally, there is no information available from available from the interferometer as to whether the phase difference is increasing or decreasing. This can be a problem, for example, in plasmas that initially increase in density, and drop before reaching steady state.

The problems of amplitude-phase ambiguity and phase uncertainty can be solved by frequency modulation (FM) of the source of a homodyne interferometer. A changing plasma density appears as a time varying phase shift in the IF signal. While effective, this approach adds to the complexity and cost of the system, and can potentially add phase noise and drifts. In addition, the frequency response of the instrument is limited by the modulation frequency ( f = 1 MHz).

A simple solution to the problems encountered in these type of homodyne instruments, without resorting to a heterodyne system, is to generate two IF outputs in phase quadrature, $VI_{F1} = A\sin(\theta)$ and $VI_{F2} = A\cos(\theta)$. One output can then be divided by the second, yielding a result containing phase information only.

$$\frac{V1_{F1}}{V1_{F2}} = \tan\theta \quad (1)$$

Although other designs are possible [3],[4], the most direct way to generate phase quadrature outputs is to use a commercially available quadrature IF mixer. A description of the implementation and operating characteristics of a 70 GHz quadrature homodyne system utilizing a packaged quadrature mixer is the subject of this paper. This system combines high sensitivity with fast time response, superior to that of an FM homodyne system, at low cost. In addition, the simplicity of the instrument provides a high level of reliability. Many interferometers are required to operate in high ambient magnetic fields (B-fields), e.g. around magnetic confinement plasma devices. Typically, microwave sources require the use of a series isolator to prevent reflected power from returning to the oscillator causing frequency shifts. Commercially available isolators are ferromagnetic devices which must be shielded from background B-fields in order to function correctly and not be damaged. Since magnetic shielding is of general interest, the design and construction of the shielding for the isolator of this system is also described in this paper.

**2. Interferometer Circuit**

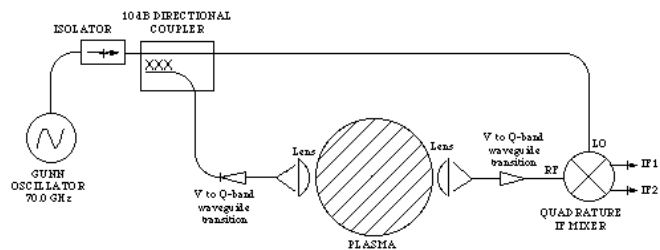

Figure 1 Interferometer setup

The interferometer circuit is shown in Fig. 1, above. The isolated Gunn oscillator output is divided by a 10 dB directional coupler with most power (90%) going to the LO input of the quadrature mixer, and the remainder directed into the plasma. The mixer is placed as physically close to the oscillator as possible in order to provide sufficient LO power, PLO. For example, the mixer used in this circuit, an Epslion-Lambda Electronics model ELMIX72, requires
PLO greater than or equal to 15 dBm, while the Gunn oscillator (Epsilon-Lambda model

ELV173) has an output power of only 17.8 dBm. Such high LO power requirements are typical of millimeter-wave quadrature mixers. The remaining source power launched into the plasma easily provides enough RF power to the mixer to give a signal to noise ratio greater than 10³ for typical lab plasma parameters, as will be shown in the next section.

In order to better collimate the microwave beam, lower band (Q-band, 33-50 GHz) horns and lenses have been used. The horns are standard high gain horns, G = 20 dB. The lenses are spherical, high density polyethylene (HDPE), and are easily and inexpensively made [5]. Phase quadrature IF outputs are generated by the quadrature mixer. Quadrature mixers in millimeter-wave ranges tend to be quite narrow band. For example, the ELMIX72 has a ± 200 MHz bandwidth, centered at 70.0 GHz. To accommodate the narrow band of the mixer the oscillator must either be matched to the mixer, or have enough tuning range, while maintaining sufficient phase stability, to tune to the mixer band. In this case, the oscillator and mixer were purchased as a matched pair.

The outputs of the quadrature mixer are $VI_{F1} = A_1\cos(\theta) + V_{OFF1}$, $VI_{F2} = A_2\sin(\theta + \theta_0) + V_{OFF2}$. $V_{OFF1}$ and $V_{OFF2}$ are DC offsets and $\theta_0$ is the quadrature phase error. These may be viewed on an oscilloscope or digitized directly. Ideally, $A_1 = A_2$, $V_{OFF1} = V_{OFF2} = 0$, and $\theta_0 = 0$, so that the outputs are exactly balanced with zero offset, and exactly in quadrature. In this case the expression for theta reduces to (1). In practice there is some amplitude imbalance, offset, and quadrature error which are functions of frequency and LO power, so that the ideal conditions stated above are not met. However, for a given frequency and LO power, the system can be calibrated by determining $A_1$, $A_2$, $V_{OFF1}$, $V_{OFF2}$ and $\theta_0$, so that theta remains the only unknown. The phase shift as a function of time, $\theta(t)$, can easily be computed from:

$$\theta(t) = a\tan\left\{\frac{1}{\cos\theta_0}\left(\frac{A_1}{A_2}\right)\left(\frac{V_{1F2}(t)-V_{off2}}{V_{1F1}(t)-V_{off1}}\right) - \sin\theta_0\right\} \qquad (2)$$

Calibration to determine A1, A2, VOFF1, VOFF2 and $\theta_0$ is accomplished by adding a variable phase shifter to the reference leg and sweeping through a $2\pi$ phase shift with no plasma present. Alternately, the receiving horn and mixer can be mounted on a moveable base, provided there is enough flexibility in the reference waveguide (in this case less than 5 mm). When the receiving horn is moved toward the launch horn to sweep through a $2\pi$ phase shift (with no plasma), the IF outputs trace out sinusoidal voltages with amplitudes $A_1$, and $A_2$, DC offsets $V_{OFF1}$, and $V_{OFF2}$, and have an error $\theta_0$ in phase quadrature. These can be measured to obtain calibration values.

The phase shift is related to the electron density by:

$$\theta(t) = \frac{\omega}{c}\int_0^L \left[1 - \sqrt{1 - \frac{\omega_{pe}^2(x,t)}{\omega^2}}\right] dx \quad ; \quad \omega_{pe}^2(x,t) \equiv \frac{n_e(x,t)e^2}{\varepsilon_0 m_e} \qquad (3)$$

where ne(x,t) is the electron density, e and me are the electron charge and mass respectively, e0 is thepermittivity of free space, and a slab plasma is assumed. The normalized density profile, $\alpha n_e(x)$, can be measured by a scanned Langmuir probe so that eq. (3) may be integrated numerically. Alternately, the chord averaged density can be computed eliminating the integral. Magnetic shielding is essential if the ambient field at the microwave isolator exceeds 20 Gauss, since it contains a magnet. The source must be kept close to the vacuum window, and therefore in the background Bfield, in order to avoid excessive power loss. The shield used here consists of two concentric iron pipes and a layer of mu-metal, as shown in the assembly drawing, Fig. 2.

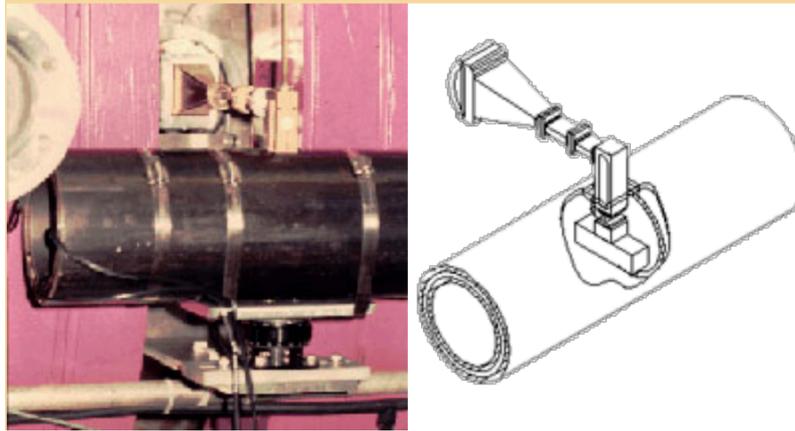

Fig. 2. Magnetic Shielding: 1) Iron tubing, 18 in. x 0.25 in. thick, 2 layers (innermost mu-metal layer not shown), 2) Gunn oscillator with heat sink, 3) Directional coupler, 4) lens.

The iron has a fairly low magnetic constant, mu = 100, but is not easily saturated. The opposite is true for the mu-metal (mu > 10,000). To prevent saturation it is placed on the innermost wall of the pipe. Such multiple shells have been shown to be highly effective as shields [6]. For example, three concentric shields each with mu = 100 are as effective as a single one with mu = 1000. The axis of the shield is parallel to the stray magnetic field from the plasma device. In our application the field at the isolator is reduced from 200 Gauss to 2 Gauss. Care must be taken to secure the iron shields lest they twist when the B-field is turned on.

3. INTERFEROMETER PERFORMANCE

Figure 3 below shows both interferometer outputs verses time, along with plasma discharge current on the **LA**rge **P**lasma **D**evice (LAPD) at UCLA [7]. Calibration by movement of the receiving horn, as discussed above, yielded A1 = 37 mV, A2 = 26 mV, $V_{OFF1}$ = 135 mV, $V_{OFF2}$ = -25 mV, and $\theta_0$ = 7 degrees. Using eq. (2) together with a normalized density profile measured by a scanned Langmuir probe, eq. (3) was integrated to give a density profile at time t = 4.5 ms as shown in figure 3(b) shown below:

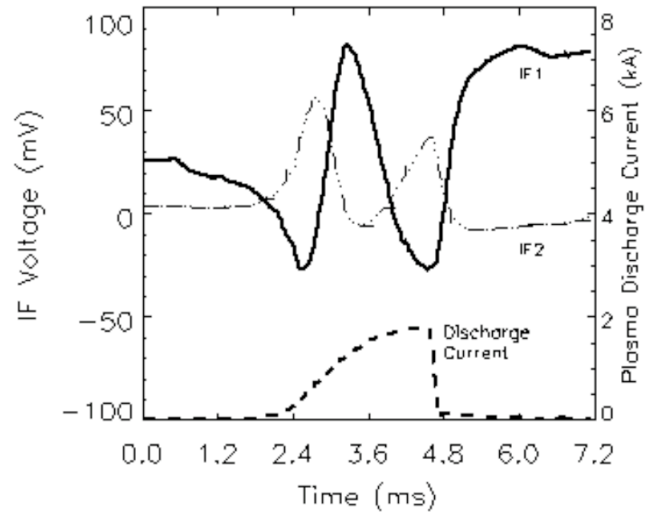

Fig. 3. (a) Interferometer outputs IF1 ($A\cos(\theta)$), IF2 ($A\sin(\theta)$), and plasma discharge current vs. time. The interferometer signal persists for about 100 ms after the discharge is terminated. The long decay time is due to the low electron temperature ($T_e < 1.5$ eV) after the discharge, and the 10 m length of the device.

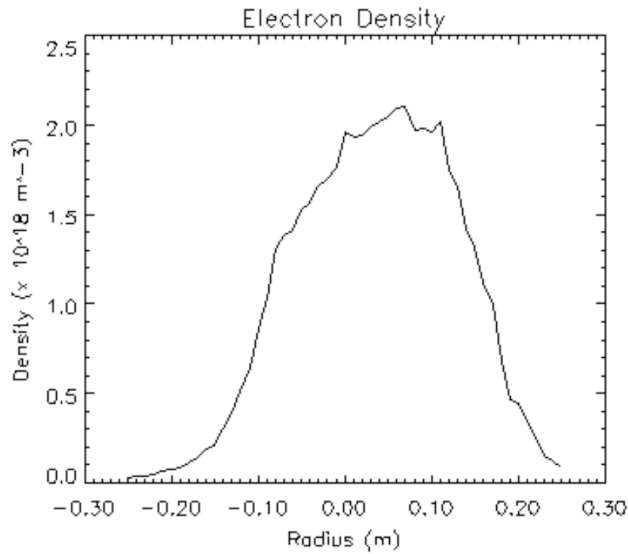

3(b) Interferometer phase vs. time, $\theta(t)$. The rapid change at t = 4.6 ms reflects the termination of the discharge

In addition the average electron density vs. time, n(t), could be computed. The frequency response of the instrument is limited only by the response of the quadrature mixer, which is DC - 100 MHz in our case. Since the interferometer responds to phase shift, the accuracy of the instrument is limited by system phase noise. Total system phase noise is the sum of phase fluctuations from the Gunn oscillator and the nonlinear phase response of the quadrature mixer (variation of A1, A2, $V_{OFF1}$, $V_{OFF2}$, and $\theta_0$ with frequency and LO power). The maximum phase noise for the Epsilon-Lambda ELV173 oscillator is of order 0.1 degree. The ELMIX72 mixer is specified to have a phase quadrature error, $\theta_0$ of order 10 degrees. While $\theta_0$ can be determined for a given frequency and LO power as discussed above, its variation with frequency and power is not specified by the manufacturer. The user must therefore assume the maximum specified quadrature error in order to determine system accuracy. Taking the maximum error in theta as the difference between $\theta_0 = 10$ degrees and $\theta_0 = 0$, using eq. (2) yields a phase error of less than 0.1 degree. The phase errors introduced by nonlinear changes in IF output amplitudes and offsets ($A_1$, $A_2$, $V_{OFF1}$, $V_{OFF2}$) of the mixer are more difficult to determine. No specifications for these variations are given by the manufacturer. However, RF to IF conversion loss is typically specified over the operating band. In the case of the ELMIX72, conversion power loss is given to vary less than 19%. Since $P = V^2/R$, the IF voltage variation into a constant impedance is therefore less than 9.1%. To determine the variations in IF offset voltages, $VI_{F1}$ and $VI_{F2}$ were measured in the 70 GHz ± 200 MHz band. It was found that $V_{OFF1}$ and $V_{OFF2}$ both had variations of approximately 5%. For phase error analysis, this was doubled to 10% to give a conservative estimate. Using these offset and amplitude variations in eq. (2), Monte Carlo analysis gives a phase error < ±1.8 degrees. Summing the phase errors due to oscillator phase noise, mixer noise, mixer quadrature error, and IF voltage amplitude and offset variations gives a total system phase noise < ±0.1 degrees (oscillator) + ±0.1 degree (quadrature error) ±1.8 degrees (nonlinear IF) < 2 degrees (total). This is a conservative estimate. In the case of the LAPD, with a plasma diameter of 0.50 m and nominal average $n_e = 1.0 \times 10^{18}$ m$^{-3}$, a two degree phase error corresponds to an error in average density of 0.6%.

The noise level of the LAPD interferometer has been measured to be $P_N < -80$ dBm, a typical level for laboratory plasma systems. The power at the receiver horn can be estimated by the radar equation for identical antennas [8] and including collisional damping by the plasma [9],

$$P_1 = P_0 \frac{G^2 c^2}{4\pi^2 l^2 f^2} e^{-\frac{1}{\delta}} \qquad (4)$$

Where G is the antenna gain, c the speed of light, l is the distance between horns, f is the microwave frequency, and $\delta$ the energy damping length. The collisional damping length is given by

$$\delta = \frac{2f^2 c}{f_{pe}^2 \nu_{ei}} \sqrt{1 - \frac{f_{pe}^2}{f^2}} \qquad (5)$$

where $f_{pe}$ is the electron plasma frequency, and $\nu_{ei}$ is the electron-ion collision frequency. $\nu_{ei} = 2.91 \times 10^{-12} Z n_e \ln(LT_e^{\frac{3}{2}})$ where Z is the ion mass ratio, $L = 12\pi n_e \lambda_D^3$, and $\lambda_D = 7.43 \times 10^{-4} \sqrt{\frac{T_e}{n_e}}$ is the Debye length [9]. $T_e$ is in electron volts. Using nominal LAPD plasma parameters, f = 70 GHz, $n_e = 10^{18}$ m$^{-3}$, $T_e = 10$ eV, and Z = 4 (Helium plasma) gives $\delta$ = 7514 m, while for $n_e = 5 \times 10^{18}$ m$^{-3}$, $T_e = 1$ eV, $\delta = 13.6$ m. With such a large value for delta, plasma damping is obviously negligible for a laboratory-size plasma, and it is reasonable to set $e^{-\frac{1}{\delta}} = 1$. Further, taking G = 20 dB, l = 1m, and $P_0$ = 0 dBm gives PR > -30 dBm. The mixer power conversion loss is roughly 10 dB. The loss in the receiver horn to mixer waveguide (V band, 3.4 m long) is < 10 dB (8 dB measured). The mixer signal power, S, is therefore S = $P_R$ - 10 dBm - 10 dBm > -50 dBm. Thus the signal to noise ratio, S/N > -50 dBm/-80 dBm = $10^3$.

## 4. CONCLUSIONS

A relatively inexpensive 70 GHz millimeter-wave interferometer has been constructed and tested on a laboratory plasma. The hardware configuration (Fig. 1) is textbook simple, and the output unambiguously yields the phase shift due to plasma in the beam path. Since the interferometer has a large bandwidth and small phase jitter, it can also be used to measure line-averaged density perturbations from plasma waves. The interferometer has been operated reliably on a daily basis, with no tuning or attention, on the LAPD device for the past 20 years.

**Acknowledgments**
This work was supported by the Department of Energy, the Office of Naval Research, and the Cal. Space Institute. The LAPD device is part of the Basic Plasma Science Facility funded by DOE and NSF.